\documentclass[aps,twocolumn,pra,showpacs,amsmath,amssymb,superscriptaddress,hyperref]{revtex4-2}
\usepackage{epsfig,graphicx}
\usepackage{bookmark}
\usepackage{physics}
\usepackage{braket}
\usepackage{xcolor}

\newcommand{\be}{\begin{equation}}
\newcommand{\e}[1]{\label{#1}\end{equation}}
\def\bea{\begin{eqnarray}}
\def\ea#1{\label{#1}\end{eqnarray}}

\def\ee{\end{equation}}
\def\eea{\end{eqnarray}}
\def\bes#1{\begin{subequations}\label{#1}}
\def\ese{\end{subequations}}

\begin{document}

\title{Consequences of the single-pair measurement of the Bell parameter}

\author{Marco Genovese}
\affiliation{INRIM - Istituto Nazionale di Ricerca Metrologica, strada delle Cacce, 91 10135 Torino, Italy.}
\affiliation{INFN, sede di Torino, via P. Giuria 1, 10125 Torino, Italy.}
\author{Fabrizio Piacentini}
\affiliation{INRIM - Istituto Nazionale di Ricerca Metrologica, strada delle Cacce, 91 10135 Torino, Italy.}

\date{\today}

\begin{abstract}
Bell inequalities represent a milestone for contemporary Physics, both for quantum foundations investigation and technological applications (e.g., quantum communication and entanglement certification).
Although loophole-free tests have been recently performed, a strong debate is still ongoing on the actual meaning of Bell inequality tests, for example on the possible additional hypotheses (end eventual loopholes) to be included in Bell's theorem, as well as on the implications for certain interpretations of quantum mechanics.
A recent work [S. Virz\`{i} et al., \emph{Quantum Sci. Technol.} \textbf{9}, 045027 (2024)] challenges some of the statements appeared in this debate, achieving for the first time an experimental estimation of the entire Bell-CHSH parameter from a single entangled pair thanks to a weak-interaction-based measurement approach.
Here we analyse the implications of this result for quantum mechanics foundations investigation, illustrating how it can tackle some of the aforementioned interpretations of Bell inequality tests and, more in general, of quantum mechanics itself.
\end{abstract}

\maketitle



\section{Introduction}

Bell inequalities \cite{b1,b2} are to be considered one of the most significant and powerful results of 20th century physics.
Despite their simplicity, they have a huge predictive power, allowing to exclude any alternative to quantum mechanics (QM) where measurement results are pre-determined by some inaccessible ``hidden'' variable(s) and locality is preserved, dubbed Local Hidden Variables Theories (LHVTs).
Thus, they have assumed a leading role in the studies on the foundations of quantum mechanics \cite{prep,stan} and, more in general, in foundations and philosophy of science \cite{l}.
Furthermore, on the applicative side, they have become a fundamental tool for quantum technologies \cite{qt1,qt2,qt3,qt4,qt5,qt6}.\\
Very shortly, Bell's theorem \cite{b2} states that, by quantifying correlations between measurements on the two components of a bipartite entangled state, there exist certain inequality that are always satisfied by any LHVT but, conversely, can be violated in a QM framework.
As an example, if one considers the correlations between Alice's and Bob's measurements performed randomly choosing between two possible settings and respectively dubbed $A_i$ and $B_j$ ($i,j=1,2$), by averaging on multiple experiments one can write the correlation parameter:
\begin{equation}\label{B}
  \mathfrak{B}= \ev{A_1 B_1+ A_1 B_2 + A_2 B_1 - A_2 B_2}.
\end{equation}
Then, it can be demonstrated that $|\mathfrak{B}|\leq2$ for every LHVT (the so-called CHSH inequality), but can reach $2 \sqrt{2}$ in QM \cite{prep}.
The hypotheses on which Bell's theorem relies are:
\begin{itemize}
  \item \emph{Realism}: there exist (inaccessible) hidden variables with a certain probability distribution, whose knowledge would allow restoring the determinism of classical physics;
  \item \emph{Independence of measurements}: Alice and Bob can realise two independent measurements, where the choices of experimental settings $i,j$ on Alice's and Bob's side are independent and space-like (non-communicating) separated events;
  \item \emph{Bell locality}: the probability of observing a measurement outcome $a_i$ on  Alice's side and $b_j$ on Bob's one factorizes in two independent probabilities, only related by the common hidden variable(s).
\end{itemize}
The path towards a conclusive test of them eliminating every possible loophole was long \cite{prep,stan,test1,test2,test3,test4,test5,test6,test7,test8}, but was eventually concluded in 2015 by three experiments \cite{lf1,lf2,lf3}.
Thus, LHVTs are excluded as a viable alternative to QM: every theory in which results are predetermined should contain some peculiar ``non-classicality'' hypothesis, as nonlocality \cite{int,nel,deb1,deb2,deb3,deb4,w}, superdeterminism \cite{sd1,sd2,sd3,sd4,sd5,sd6,sd7,sd8,sd9}, wormholes at small scales (ER=EPR \cite{mal}), extra dimensions \cite{dim1,dim2,dim3}, etc.\\
All the Bell inequality tests realised so far required experimental setups allowing Alice and Bob to randomly switch, respectively, between the $A_1,A_2$ and $B_1,B_2$ observables measurement, because of their incompatibility (non-commutativity).
This implies that each copy of the bipartite entangled state exploited actually contributes to estimate just one of the four terms of the Bell-CHSH parameter $\mathfrak{B}$ in Eq. \eqref{B}, since the quantification of the entire parameter is forbidden by the impossibility of measuring, at once, all the observables needed for such a task.\\
Nevertheless, a recent work \cite{wchsh} relaxed this request, by implementing a sequence of weak measurements (WMs) \cite{wm1,wm2,wm3,swm1,swm2,swm3} on both Alice's and Bob's side.
This result was the final achievement of a line of research pursued by the quantum foundations community where WMs were applied to an entangled state in different ways, e.g. for testing Leggett-Garg inequalities \cite{lg1,lg2,lg3,lg4,lg5,lg6,lg7,lg8}, by implementing WMs on just one of the entangled particles \cite{e1,e2}, or realizing for both particles a weak measurement followed by a projective one \cite{e3}.
The aforementioned experiment \cite{wchsh} demonstrated the unprecedented possibility of estimating the \emph{entire} Bell-CHSH parameter from each entangled photon pair by having both photons undergo a sequence of two WMs of their polarisation.
Furthermore, quantum tomographic reconstruction of the entangled state before and after the WMs highlighted how the entanglement only suffered minor decoherence from the (weak) polarisation measurement process, hence allowing entanglement certification without sacrificing any entangled pair produced and, as a consequence, increasing the quantum resources available for further quantum technology protocols.\\
Although the experiment was not planned to be loophole-free, conceptually nothing forbids to realise a loophole-free version of it.
The detection loophole can be eliminated with low-absorbing optical components and high-efficiency single-photon detectors (some of them already available on the market).
Also, measurement independence can be realised when Alice a Bob, following independent random number generators, respectively choose their polarisation measurements settings to be $[\theta_1^{(A)},\theta_2^{(A)}]=[0, 3\pi/4] \vee [\pi/2, 3\pi/4]$ and $[\theta_1^{(B)},\theta_2^{(B)}]=[-3\pi/8, 3\pi/8] \vee [\pi/8, -\pi/8]$, being $\theta_i^{(K)}$ ($i=1,2;\;K=A,B$) a rotation angle with respect the $\{H,V\}$ measurement basis ($H$ and $V$ indicating the horizontal and vertical polarizations, respectively).
Such a framework leads to $\mathfrak{B}=2\sqrt{2}$ when both Alice and Bob choose the first or second pair of settings, and to $\mathfrak{B}=-2\sqrt{2}$ in case of crossed choice, anyway achieving a maximal violation of the CHSH inequality.

This result represents a very significant progress in the studies in this field, providing on the one hand a new tool for testing resources (entangled states) to be used in quantum technologies without substantially deteriorating them, and, on the other hand, allowing to deepen several arguments related to understanding entanglement.\\
In Ref.\cite{relindexp}, a modified version of this experiment allowed to quantitatively observe the interplay among local and non-local correlations and their relation with the uncertainty principle, also studied in \cite{relind,jpc}.
In particular, in these last references the violation of Bell inequalities was conjectured as stemming from the fact that ``the outcomes of incompatible measurements cannot be described by a joint non-contextual reality''.
Notwithstanding the relevance of the use of incompatible measurements and the capital role of uncertainty principle in discussing the correlations leading to Bell's theorem violation, the discussion of these last two references should be remodulated at the light of Ref. \cite{wchsh}.\\
In the following, we present several implications of this experiment on the foundations of quantum mechanics.
In particular, we will demonstrate that it casts serious concerns on certain interpretations of QM, while its results remain somehow unclear (at the interpretation level) for other theories and models (not analysed here, like, e.g., the two-state-vector formalism \cite{TSVF}).

\section{Impact on counterfactual statements}

Beyond the Bell's theorem hypotheses listed above, there was a wide and long debate \cite{cdf1,cdf2,cdf3,cdf4,cdf5,cdf6,cdf7,cdf8,cdf9,cdf10,cdf11} if a further hypothesis is needed, i.e. the one concerning \emph{counterfactual definiteness}, namely the ability to speak ``meaningfully'' of the definiteness of the results of measurements that have not been performed, since the joint measurements of incompatible observables appearing in Eq.\ref{B} must be performed, in the traditional (projective) quantum measurement framework, in independent experiments on identical copies of the same entangled state.
This argument was reconsidered even very recently in Ref. \cite{aie}, stating that ``This assumption is counterfactual because the results of distinct experiments must be described by different random variables'', to claim that: ``Bell's theorem supposedly demonstrates an irreconcilable conflict between quantum mechanics and local, realistic hidden variable theories. In this paper we show that all experiments that aim to prove Bell's theorem do not actually achieve this goal''.\\
Eventually, it seems that counterfactual definiteness is not really needed as a further hypothesis at least for some versions of Bell theorem, as discussed in \cite{lam1,lam2}.
Nevertheless, Ref. \cite{lam2} reports that ``Since it is not possible to perform more than one measurement on a particle without disturbing it, (in the case of photons, the particles are annihilated after they are absorbed) the most popular solution is to interpret \cite{Stapp} according to the CFD recipe...''.
Of course, the aforementioned experiment \cite{wchsh} completely challenges this statement, eliminating once and for all any discussion about counterfactual statements related to Bell inequalities.\\
For instance, Stapp's proof of Bell inequalities built to exclude counterfactual hypotheses \cite{Stapp} must be completely reconsidered at the light of these results, since it is removed the need for any counterfactual statement for testing the CHSH inequality (and, in principle, other Bells inequalities, like Wigner's one \cite{wign1,wign2}).
The same holds for the recent discussion appearing in Ref. \cite{hnilo}, stating that: ``It is impossible measuring with different angle settings at the same time slot […] Sica's approach also makes easily visible the unavoidable limitation that measuring with different settings requires measuring at different times. This limitation implies that Bell inequalities are valid only if arbitrary time ordering of the series is assumed always possible.''
The elimination of any counterfactual statement related to Bell's inequalities represents a key contribution to the studies on the foundations of quantum mechanics, but several other significant impacts follow from this achievement.

\section{Considerations about probability spaces}

One relevant consequence of Bell inequalities is that, according to Fine's theorem \cite{fine}, their violation implies that do not exist marginal pair probabilities $p(a_i,b_j)$ (i,j=1,2) justifying this violation, and attempts to reconstruct them lead, for instance, to the emergence of ``negative probabilities'' \cite{np}.
This observation was then developed in elaborating non-Kolmogorovian probabilistic interpretations \cite{cdf9,nonKolm1,nonKolm2,nonKolm3,nonKolm4,nonKolm5,nonKolm6} of Bell inequality tests, asserting that these tests incorrectly put statistical data belonging different (incompatible) sampling experiments into one single probability space.\\
Without entering into the rather wide discussion on these proposals, we underline that the result of Ref. \cite{wchsh} demonstrates that it is possible to measure the Bell parameter in a single experimental setting, overcoming the objection at the basis of this Bell test interpretation line. 
The result of Ref.\cite{wchsh} does not challenge the Fine's theorem, but it shows that, although requiring measurement of incompatible observables, a ``single-shot'' measurement of the Bell parameter, i.e. the estimation of the entire $\mathfrak{B}$ quantity on each entangled pair, can be achieved via WM approach.
Of course, due to the uncertainty principle, this single-shot estimate is affected by a large uncertainty, that can be strongly reduced by averaging on repeated experiments on identically-generated entangled pairs (as for every experimental procedure).\\
A bit on the same line, in Ref. \cite{kh} it was considered the Landau identity:
\begin{equation}\label{B2}
  \mathcal{B}^2=\frac{\mathfrak{B}^2}{4}= 1-\frac14 [ A_1 , A_2 ]\cdot[B_1 , B_2]\;.
\end{equation}
If at least one of the two commutators $[ A_1 , A_2 ]$ and $[B_1 , B_2]$ is found to be zero, then one can derive, by means of quantum formalism, the inequality $|\langle  \mathcal{B} \rangle |\leq 1$, formally identical to the CHSH inequality (although with a different interpretation) and dubbed \emph{quantum CHSH inequality}.
This observation led the author to claim that the violation of Bell inequality derives from the local incompatibility of observables in QM ($[ A_1 , A_2 ]\neq0$ and $[B_1 , B_2]\neq0$), and that non-locality is not needed.
Although Eq. \eqref{B2} points out that incompatible observables are required for violating Bell inequalities, in our opinion the realisation of a CHSH inequality test where these observables are measured altogether casts serious questions on interpreting this request as the crucial physical point behind Bell inequalities, and on the fact that they ``are tests on violation of the Bohr complementarity principle'' \cite{kh}.\\
The same argument applies to the attempt to violate Bell inequalities through non-ergodicity \cite{nonerg}, since such violation is again obtained by strictly considering as disjoined the measurements pertaining to incompatible observables on both Alice's and Bob's sides (see section 9 of \cite{nonerg}).

\section{Impact on modal interpretations}

The results presented in Ref. \cite{wchsh} also challenge some interpretations of quantum mechanics, as the so-called ``modal interpretations'' \cite{modal1,modal2,modal3,modal4,d2007,stan2}, attempting to build QM as an objective, observer-independent description of physical reality.
This is achieved by stating that the relation between the formalism of quantum theory and physical reality is to be taken as probabilistic, i.e., the quantum formalism does not describe what actually happens in the physical world, but rather provides us with a list of possibilities and their probabilities: ``the state in Hilbert space is about possibilities, about what may be the case, about modalities'' \cite{d2007}.\\
In summary, here the mathematical expression of the quantum state represents situations with definite physical properties, even if this state is a superposition of eigenstates of the corresponding observables.
These properties are assigned depending on the context, including the measurement apparatus and settings (for example, on the choice of measuring the spin along the $x$ or $z$ axis), and are defined in a bi-orthogonal decomposition, including the representation of the state in the basis of the eigenvectors of the measured observables and the corresponding pointer states.
The assignment of an objective property to a quantum system, related to the choice of the observable measured by an experimental apparatus, can hardly agree with the possibility of measuring incompatible observables on the same state (of course obeying the Heisenberg uncertainty principle, like it occurs with WMs due to the large measurement uncertainty).
The results in Ref. \cite{wchsh} contradict this theoretical framework, in which the Bell parameter measurement should be related to the context of the setting (i.e., incompatible observable to be measured) chosen in Alice's and Bob's measurement apparatuses.\\
This is highlighted in a recent formulation \cite{gran} of this kind of interpretations, mainly relying on the fact that ``the extended commutative algebra of \emph{quantum and classical measuring apparatus states} is the fundamental mathematical object to completely describe a system within a context''.
In this framework, the author states that:\\
``Another noteworthy remark is that by completing QM in this more general framework, Bell's inequalities cannot be written anymore, because they involve a counterfactual mix up of different classical contexts.
In a more explicit way, let us consider a standard Bell-CHSH situation, with two entangled particles a and b, and measurements of $A_1$ or $A_2$ for a, $B_1$ or $B_2$ for b.
One has $[A_m,B_j] = 0$ for $m,j = 1,2$, and in order to get a violation of Bell's inequalities both commutators $[A_1,A_2]$ and $[B_1,B_2]$ must be nonzero.
In a more general way, both algebras ${A_m}$ and ${B_j}$ must be non-commutative.
But if one considers the extended algebras for a and b, which are commutative and sectorized, there is no more violation in this extended algebra.
In some sense, the violation of Bell-CHSH inequalities is never `measured', but inferred assuming that the statistics are determined by the properties of the system (the entangled pair) in a context free way.''\\
It is evident that, in this interpretation, it is impossible to describe a scenario in which incompatible variables are measured on the same quantum state as in Ref. \cite{wchsh}, since they will belong to different ``contexts''.
This means that the measurement of the entire Bell-CHSH parameter on a single entangled pair strongly challenges the possibility of building a valid QM interpretation based on ``modalities''.\\
Similar arguments pertain Kupczynski's (again context-dependent) interpretation \cite{nonKolm6,k}.
For example, in \cite{k} it is stated, discussing Bell inequalities tests, that: ``Since variables $(A,A')$ as well as $(B,B')$ cannot be measured jointly, neither Nx4 spreadsheets nor a joint probability distribution of $(A,A',B,B')$ exist, thus Bell CHSH inequalities may not be derived''.
Again, results of Ref. \cite{wchsh} are in contrast with these statements.

\section{Conclusions}

We analysed the impact and implications on QM foundations of a recent experiment able to realize an entanglement-preserving, single-pair measurement of the Bell-CHSH parameter \cite{wchsh}.
This analysis demonstrated how this experimental achievement challenges several conclusions and discussions on the meaning of Bell inequalities as well as certain QM interpretations, highlighting how this result represents, in our opinion, one of the most fruitful and innovative achievements in quantum foundations in the recent years.

\section{ACKNOWLEDGEMENTS}

This work was financially supported by the project Qutenoise (call ``Trapezio'' of Fondazione San Paolo).

\end{document}